\documentclass[final,3p,times,twocolumn,sort&compress]{elsarticle} 

\journal{Physics Open}

\usepackage{bm}
\usepackage{amsmath}
\usepackage{bm}
\usepackage{xspace}

\usepackage{mwe}
\usepackage{widetext}

\usepackage{graphicx}

\usepackage{natbib}

\usepackage[separate-uncertainty]{siunitx}
\usepackage{amsmath, amssymb, commath, mathtools}
\usepackage{physics}

\newcommand*\erfc[1]{\mathrm{erfc}\left( #1 \right)}
\renewcommand*\erf[1]{\mathrm{erf}\left( #1 \right)}
\renewcommand*\exp[1]{\mathrm{exp} \left( #1 \right)}
\newcommand*\expo[1]{\mathrm{e}^{#1}}
\renewcommand*\ln[1]{\mathrm{ln} \left( #1 \right)}

\begin{document}

\begin{frontmatter}



\title{Solving a nonlinear analytical model for bosonic equilibration}


\author{N. Rasch and G. Wolschin\corref{cor}}
\ead{g.wolschin@thphys.uni-heidelberg.de}

\address{Institut f{\"ur} Theoretische Physik der Universit{\"a}t Heidelberg, Philosophenweg 12-16, D-69120 Heidelberg, Germany, EU}

\cortext[cor]{Corresponding author}

\begin{abstract}
An integrable nonlinear model for the time-dependent equilibration of a bosonic system that has been devised earlier  
is solved exactly with boundary conditions that are appropriate for a truncated Bose-Einstein distribution, and include the singularity at $\epsilon = \mu$.
The buildup of a thermal tail during evaporative cooling, as well as the transition to the condensed state are accounted for. 
To enforce particle-number conservation during the cooling process with an energy-dependent density of states
for a three-dimensional thermal cloud, a time-dependent chemical potential is introduced.
\end{abstract}


\end{frontmatter}
\newpage

\section{Introduction}
The formation of a Bose-Einstein condensate (BEC) from a cloud of bosonic
atoms in a trap represents a phase transition from a normal phase 
with a small but finite negative chemical potential $\mu<0$ to a condensed phase
with $\mu\rightarrow 0$ \cite{an95,ket95,dal99}. To achieve such a phase transition, successive evaporative cooling is used that removes high-velocity atoms. When equilibrating, the number of condensed particles rises due to a transfer from the nonlinear kinetic region into the coherent region \cite{svi91,kss92,kas97}, and an isotropic thermal tail \cite{an95,BookPitaevskii} develops in the ultraviolet, smearing out the sharp cut at higher energies that corresponds to evaporative cooling.

In this work, we investigate an analytical model for the time-dependent equilibration in a gas of cold bosonic atoms. It is  based on a
nonlinear boson diffusion equation (NBDE) that has been proposed in Refs.\,\cite{gw18,gw18a}.
We solve it exactly with initial conditions that are suitable for evaporative cooling, plus boundary conditions
that are appropriate for a truncated Bose-Einstein distribution.
With the additional requirement of particle-number conservation,
and the energy-dependent density of states for a three-dimensional isotropic cloud we calculate the time-dependent
number of particles in the thermal cloud, and in the condensate, thus accounting for the time-dependent
condensation and thermalization.

Analytical solutions for physically meaningful nonlinear partial differential equations
are of great interest in several fields of physics, but are rarely available.
In case of the nonlinear boson diffusion equation \cite{gw18,gw18a} which we consider in this work with initial and boundary conditions,
no exact solutions in $2+1$ and higher dimensions are presently known.
We discuss it here in $1+1$ dimensions (energy and time), which is appropriate for a condensing isotropic three-dimensional thermal cloud of cold atoms. 

In Sec.\,2, the basic nonlinear equation is briefly reviewed, and its exact solutions with initial and boundary condition
are outlined. Various special cases are investigated in detail in Sec.\,3. Starting with a fixed temperature $T$ and variable chemical potential, we proceed to the solution for
fixed chemical potential $\mu=\,$const with corresponding boundary conditions, and analyze and compare the two special solutions. To schematically account for evaporative cooling,
the time-dependent solutions for differing initial and final temperatures are then examined in Sec.\,4. The time-dependent partition function
is derived for a single cooling step from a temperature $T_\text{i}$ to $T_\text{f}$, the occupation-number distribution is obtained analytically from the
partition function, and it is shown to have the correct asymptotic behaviour. In Sec.\,5, we implement overall particle-number conservation into
the model, using the density of states for a three-dimensional isotropic Bose gas. The conclusions are drawn in Sec.\,6.
\section{Nonlinear Boson Diffusion Equation and solutions with boundary conditions}
\label{sec:solutionofnonlinearPDE}
The kinetic equation for the energy-dependent
single-particle occupation numbers $n_j\equiv\,n\,(\epsilon_j,t)$ in a spatially uniform Bose system with isotropic momentum distribution can be written as
\begin{eqnarray}
\frac{\partial n_1}{\partial t}=\sum_{\epsilon_2,\epsilon_3,\epsilon_4}^\infty  \langle V_{12,34}^2\rangle\,G\,(\epsilon_1+\epsilon_2, \epsilon_3+\epsilon_4)\times\qquad\qquad\\ \nonumber
\bigl[(1+n_1)(1+n_2)\,n_3\,n_4-
(1+n_3)(1+n_4)\,n_1\,n_2\bigr]\,.
 \label{boltzmann}
\end{eqnarray}
Here,  $\langle V^2\rangle$ is the second moment of the interaction and $G$ the energy-conserving function. The collision term can also be converted
to the form of a master
equation with gain and loss terms, respectively \cite{gw18,gw18a}
\begin{equation}
\frac{\partial n_1}{\partial t}=(1+n_1)\sum_{\epsilon_4} W_{4\rightarrow1}\,n_4-n_1\sum_{\epsilon_4}W_{1\rightarrow4}(1+n_4) 
 \label{boltz}
\end{equation}
with the transition probability
\begin{equation}
W_{4\rightarrow1}=\sum_{\epsilon_2,\epsilon_3}\, \langle V_{12,34}^2\rangle\,G\,(\epsilon_1+\epsilon_2, \epsilon_3+\epsilon_4)\,(1+n_2)\,n_3
 \label{trans}
\end{equation}
and $W_{1\rightarrow 4}$ accordingly. As detailed in Ref.\,\cite{gw18a}, the summations are then replaced by integrations, introducing the densities of states $g_j\equiv g(\epsilon_j)$.
An approximation to Eq.\,(\ref{boltz}) can then be obtained through a Taylor expansion
around $\epsilon_4=\epsilon_1$ to second order. By introducing transport coefficients via moments of the transition probability
\begin{eqnarray}
D=\frac{1}{2}\,g_1\int_0^\infty W(\epsilon_1,x)\,x^2\dd{x},~~
 \varv=g_1^{-1}\frac{\mathrm{d}}{\mathrm{d}\epsilon_1}(g_1D)
 \label{moments}
\end{eqnarray}
one arrives at a nonlinear partial differential equation for $n\equiv n\,(\epsilon,t)\equiv n\,(\epsilon_1,t)$ \cite{gw18,gw18a}
 \begin{equation}
\frac{\partial n}{\partial t}=-\frac{\partial}{\partial\epsilon}\Bigl[\varv\,n\,(1+n)-n^2\frac{\partial D}{\partial \epsilon}\Bigr]+\frac{\partial^2}{\partial\epsilon^2}\Bigl[D\,n\Bigr]\,.
 \label{boseq}
\end{equation}
Dissipative effects are expressed through the drift term $v(\epsilon,t)$, diffusive effects through the diffusion term $D(\epsilon,t)$.
In the limit of constant transport coefficients, the nonlinear boson diffusion equation (NBDE)  
for the occupation-number distribution $n(\epsilon,t)$
becomes
\begin{equation}
\frac{\partial n}{\partial t}=-\varv\,\frac{\partial}{\partial\epsilon}\bigl[n\,(1+n)\bigr]+D\,\frac{\partial^2n}{\partial\epsilon^2}\,.
 \label{bose}
\end{equation}
It contains the nonlinearity of the system in an essential way, although not fully due to the assumption of constant transport coefficients.
We solve this kinetic equation exactly with suitable initial conditions for cold quantum gases similar to Ref.\,\cite{gw18a}. However, we now 
include the singularity at $\epsilon=\mu$ in the initial conditions, and later also
consider appropriate boundary conditions -- in both cases with the aim to derive analytical solutions that converge toward the proper 
Bose-Einstein equilibrium solution. 
It was observed already in Ref.\,\cite{gw18a} that the thermal equilibrium distribution is indeed a stationary solution of the NBDE, 
\begin{equation}
n_\text{eq}(\epsilon)=\frac{1}{e^{(\epsilon-\mu)/T}-1}
 \label{Bose-Einstein}
\end{equation}
with the chemical potential $\mu<0$, and the temperature $T=-D/\varv$ in a finite boson system.

As discussed in Refs.\,\cite{gw18,gw18a}, one of us has proposed two different approaches to solve the time-dependent equation analytically.
The first method, the nonlinear transformation
\begin{align}
    n(\epsilon,t) = -\frac{D}{\varv} \frac{\partial}{\partial\epsilon} \ln{P(\epsilon,t)},
\end{align}
reduces the NDBE to a linear Fokker-Planck equation for $P(\epsilon,t)$ with constant coefficients. Its solution can be retransformed
to obtain the occupation-number distribution, as we did in Ref.\,\cite{bgw19} for the analogous case of equilibration in a fermionic system.
Here we focus on the second approach \cite{gw18a}, where the linear transformation
\begin{align}
    n(\epsilon,t) = \frac{1}{2 \varv} w(\epsilon,t)-\frac{1}{2}
    \label{eq:noutofw}
\end{align}
is applied. The result is Burgers' equation \cite{bur48}
\begin{align}
    \frac{\partial w}{\partial t} +w \frac{\partial w}{\partial\epsilon} = D \frac{\partial^2 w}{\partial\epsilon^2}\,,
\end{align}
which can be solved by Hopf's transformation \cite{ho50},
\begin{align}
    w(\epsilon,t) = - 2 D \frac{\partial}{\partial\epsilon} \ln{ \mathcal{Z}(\epsilon,t)}\,.
    \label{eq:HopfTrafo}
\end{align}
The function \(\mathcal{Z} (\epsilon , t)\) can be interpreted as a time-dependent partition function of the bosonic system, which will be the central quantity to be computed in the following sections. This results in
\begin{align}
    \frac{\partial}{\partial t}{\mathcal{Z}}(\epsilon,t) = D \frac{\partial^2}{\partial\epsilon^2}{\mathcal{Z}}(\epsilon,t)\,
    \label{eq:diffusionequation}
\end{align}
which is a linear diffusion (or heat) equation. For an initial distribution \( n_{\mathrm{i}}(\epsilon) \), the solution for the time-dependent occupation-number distribution becomes
\begin{align}
    n(\epsilon,t) = -\frac{D}{\varv} \frac{\partial}{\partial\epsilon}\ln{\mathcal{Z}} -\frac{1}{2}
    \label{eq:Nformula} 
    \end{align}
    with
    \begin{align}
    \mathcal{Z}(\epsilon,t) = \int_{-\infty}^{+\infty} G(\epsilon,x,t)\,F(x)\,\dd{x}\,.
    \label{eq:partitionfunctionZ}
    \end{align}
No boundary conditions are required at this stage. Green's function of Eq.\,(\ref{eq:diffusionequation}) is \(G(\epsilon , x , t)\), which is simply a Gaussian
\begin{align}
    G(\epsilon,x,t)=\exp{- \frac{(\epsilon-x)^2}{4Dt}}\,.
    \label{eq:Greensnonfixed}
\end{align}
The initial condition for the diffusion equation is \(F(x)\), which depends on the initial occupation-number distribution $n_\mathrm{i}(y)$,
\begin{align}
    F(x) = \exp{ -\frac{1}{2D}\left( \varv x+2\varv \int_0^x n_{\mathrm{i}}(y) \dd{y} \right) }\,.
       \label{ini}
\end{align}
In the solution Eq.\,\eqref{eq:partitionfunctionZ} of the diffusion equation, a time-dependent prefactor should arise if Eq.\,\eqref{eq:diffusionequation} is solved exactly. We neglect this term throughout this work because it drops out when taking the logarithmic derivative.

\begin{figure}[t!]
    \centering
    \includegraphics[width = 0.48\textwidth]{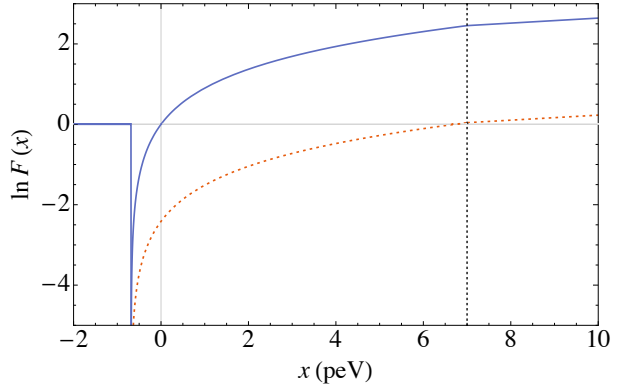}
    \caption{The argument $\ln{F(x)}$ of the exponential function in $F(x)$ of Eq.\,(\ref{ini}) (solid curve) with a singularity at  \(x = \mu\) in the definite integral of an initial distribution $n_\text{i}(y)$ given by a Bose-Einstein distribution that is truncated at $y = 7$ peV (dotted vertical line). The parameters are as in Ref.\,\cite{gw18a}, and in Sec.\,III\,C. The dashed curve is the argument 
    $\ln{F(x)}$ for the corresponding indefinite integral of the initial distribution (primitive) $A_\text{i}(x)$ with an integration constant $c\equiv 0$, see text. } 
    \label{fig0}
\end{figure}

If the initial distribution function $n_{\mathrm{i}}(y)$ has a singularity,
infinities occur in the calculation of \(\mathcal{Z} (\epsilon , t)\). In Ref.\,\cite{gw18a}, these were circumvented by confining the initial conditions to $\epsilon\ge 0$, cf. Eq.\,(20) there. In this work, we include the negative-energy region in the initial distribution. However, $F(x)$ as well as 
$\mathcal{Z}(\epsilon,t)$ and $n(\epsilon.t)$  diverge when the integral encounters a pole, as is the case for a truncated Bose-Einstein distribution at $\epsilon=\mu\le 0$. 
This is illustrated in Fig.\,\ref{fig0}, where the argument of the exponential function in $F(x)$ is shown for such an initial distribution with the parameter set of Ref.\,\cite{gw18a} that we shall also use in this work (see Sec.\,III\,C).  

We now explicitly treat such infinities by observing that the starting point of the integral in Eq.\,(\ref{ini}) -- and thus, in the expression for $   \mathcal{Z}(\epsilon , t)$ -- occurs in the nominator as well as in the  denominator of Eq.\,(\ref{eq:Nformula}) for the occupation-number distribution function due to the logarithmic derivative and therefore, it will drop out in the final result. Hence, we can use in Eq.\,(\ref{ini}) the primitive of the initial distribution \(A_{\mathrm{i}}(x)\) with the defining property \( \partial_{x} A_{\mathrm{i}} (x) = n_{\mathrm{i}} (x) \) instead of the definite integral, and \(F(x)\) can be rewritten in the form
\begin{align}
    F(x) = \exp{-\frac{1}{2D}\left( \varv x+2\varv A_{\mathrm{i}}(x) \right)}\,.
    \label{eq:F(x)}
\end{align}
With this replacement, the analytical solution of the nonlinear boson diffusion equation is still exact. In particular, for initial conditions that are confined to $\epsilon\ge 0$, the result is identical to the one obtained in Ref.\,\cite{gw18a}. However, the inclusion of a singularity at negative energies in the integral Eq.\,\eqref{eq:partitionfunctionZ} from $-\infty$ to $+\infty$ now becomes possible by integrating analytically across the pole.

There is yet no reasonable physical interpretation of the negative-energy region below the singularity at the chemical potential available, so that we confine the validity of the analytical solutions to the energy region above the chemical potential. When evaluating these solutions, it turns out that the singularity migrates in the course of the time evolution, such that the chemical potential changes, see Sec.\,\ref{sec:SolutionforTruncEquilDistr}\,A. Hence,  particle-number conservation is violated, and prohibits a physical interpretation.

Therefore, we subsequently solve the problem in Sec.\,\ref{subsec:FixedchemPot}\, by imposing a boundary condition $n(\epsilon=\mu,t)=\infty$ at the singularity $\epsilon=\mu$,  thus treating the chemical potential as a fixed parameter. With \(\lim_{\epsilon \downarrow \mu} n(\epsilon,t) = \infty\) \,$\forall$ \(t\), we have \( \mathcal{Z} (\mu,t) = 0\) from Eq.\,\eqref{eq:noutofw} and Eq.\,\eqref{eq:HopfTrafo}. Given this restriction, the diffusion equation Eq.\,\eqref{eq:diffusionequation} must be solved with one static zero point in the partition function. This requires a new Green's function 
that equals zero at \(\epsilon = \mu\) $\forall \,t$. It can be written as
\begin{align}
    \Tilde{G} (\epsilon,x,t) = G(\epsilon - \mu,x,t) - G(\epsilon - \mu,-x,t)\,,
    \label{eq:newGreens}
\end{align}
and the new partition function with this boundary condition can then be defined as
\begin{align}
    \Tilde{\mathcal{Z}} (\epsilon,t) = \int_0^\infty \Tilde{G} (\epsilon, x, t) F(x+\mu) \dd{x}\,.
    \label{eq:newformulaforZ}
\end{align}
Here, the new Green's function Eq.\,\eqref{eq:newGreens} restricts the partition function $\Tilde{\mathcal{Z}}(\epsilon,t)$ to energies $\epsilon \ge \mu$. 
With Eq.\,\eqref{eq:Nformula} and $\mathcal{Z}\rightarrow\Tilde{\mathcal{Z}}$, we obtain an expression for the occupation-number distribution with boundary conditions.
This solution for fixed chemical potential is equivalent to imposing a point symmetry around \(\mu\) in the initial distribution $n_\text{i}(\epsilon)$ that appears in \(F(x)\). 
\section{Solution for a truncated equilibrium distribution}
\label{sec:SolutionforTruncEquilDistr}
The two solution formulas in Sec.\,\ref{sec:solutionofnonlinearPDE} without and with boundary conditions at the singularity are now used to calculate explicit solutions of 
Eq.\,\eqref{bose}. Divergences within the partition functions are avoided by neglecting irrelevant factors and converging solutions are obtained. This improves the treatment in Ref.\,\cite{gw18a}, where a condition had to be introduced to complete the equilibration. For fermions \cite{bgw19} and for bosons, distributed according to a box function \cite{gw18}, converging analytic solutions have already been derived. The latter one did, however, not converge to a Bose-Einstein distribution because the singularity had been excluded. This singularity will now be considered in the initial distribution and thereby generate solutions that converge to a Bose-Einstein distribution.
\subsection{Variable chemical potential}
\label{subsec:VariableChemPot}
In order to eventually describe the process of evaporative cooling in a schematic fashion with a single cooling step through a solution of Eq.\,\eqref{eq:partitionfunctionZ} including boundary conditions, a truncated Bose-Einstein distribution with equilibrium temperature \(T=-\frac{D}{\varv}\) is chosen. This accounts for the removal of bosons above a cutoff energy \(\epsilon_i\) in the cooling process. The time-dependent occupation-number distribution should then represent the equilibration from the truncated nonequilibrium distribution
\begin{align}
    n_{\mathrm{i}}(\epsilon) = \frac{1}{\exp{\frac{\epsilon-\mu}{T}}-1} \,\Theta(\epsilon_i-\epsilon)
    \label{eq:initialdistribution}
\end{align}
to a regular Bose-Einstein distribution. The truncation at \(\epsilon_i\) is introduced through a Heaviside function \(\Theta(x)\). Different from the treatment in Ref.\,\cite{gw18a}, the initial distribution is not confined to $\epsilon \ge 0$. This has important consequences for the convergence of the solution. By choosing the initial temperature \(T\) of the truncated Bose distribution identical to the final temperature, it is apparent that no cooling process will yet occur, although equilibration takes place. The temperature of the equilibrium solution of Eq.\,\eqref{bose} is set by the transport coefficients and does not depend on the initial temperature. 

Later in Sec.\,\ref{sec:EquilforArbitraryInitialTemp}, arbitrary initial temperatures will be discussed and proper cooling will be described, but at first it is useful to examine this simple type of initial condition. The major task in the following calculations will be the evaluation of the partition function \(\mathcal{Z}(\epsilon ,t )\).
The primitive \(A_{\mathrm{i}} (x)\) of \(n_{\mathrm{i}}\) is determined up to an additional constant. This constant is neglected because of the logarithmic derivative in Eq.\,\eqref{eq:Nformula} to obtain
\begin{align}
    A_{\mathrm{i}}(x) = \begin{cases}
    T \, \ln{z^{-1}- \exp{-\frac{x}{T}}}&  x<\epsilon_i\\
    T \, \ln{z^{-1}- \exp{-\frac{\epsilon_i}{T}}}&  x>\epsilon_i \end{cases}
    \label{eq:antiderivative}
\end{align}
with the fugacity \(z \equiv \exp{ \frac{\mu}{T}}\). Already in this first step it is evident why the choice of an initial distribution at the equilibrium temperature simplifies the calculations. In Eq.\,\eqref{eq:F(x)}, the prefactor \(T\) of the primitive cancels and enables the contraction of the exponential and the logarithmic function. This step will cause difficulties when considering an arbitrary initial temperature that differs from the final (equilibrium) temperature, in Sec.\,\ref{sec:EquilforArbitraryInitialTemp}. 

Inserting \(A_{\mathrm{i}} (x)\) into Eq.\,\eqref{eq:F(x)} yields
\begin{align}
    F(x)&=\begin{cases}
    \exp{-\frac{\varv x}{2D}+\ln{z^{-1}- \exp{\frac{\varv x}{D}}}} & x<\epsilon_i \\
    \exp{-\frac{\varv x}{2D}+\ln{z^{-1}- \exp{\frac{\varv\epsilon_i}{D}}}} & x>\epsilon_i
    \end{cases} \notag \\
    &= \begin{cases}
    z^{-1}\exp{-\frac{\varv x}{2D}}-\exp{\frac{\varv x}{2D}} &  x<\epsilon_i \\
    z^{-1}\exp{-\frac{\varv x}{2D}}-\exp{-\frac{\varv x}{2D}+\frac{\varv\epsilon_i}{D}} &  x>\epsilon_i\,.
    \end{cases}
    \label{eq:F(x)calculated}
\end{align}
With Green's function Eq.\,\eqref{eq:Greensnonfixed} for variable chemical potential, the partition function \(\mathcal{Z} (\epsilon ,t )\) is evaluated. Due to the different definitions of 
\(F(x)\) in different energy regimes the integral has to be split, 
\begin{align}
    \mathcal{Z} ={}& z^{-1} \int_{-\infty}^\infty \exp{-\frac{(\epsilon-x)^2}{4Dt} - \frac{\varv x}{2D}} \dd{x} \\ \nonumber
    &- \int_{-\infty}^{\epsilon_i} \exp{-\frac{(\epsilon-x)^2}{4Dt} + \frac{\varv x}{2D}} \dd{x}\\ \nonumber
     &-\int_{\epsilon_i}^\infty \exp{-\frac{(\epsilon-x)^2}{4Dt}- \frac{\varv x}{2D} + \frac{\varv\epsilon_i}{D}} \dd{x} \notag \\ \nonumber
    ={}& \sqrt{\pi Dt} \, \exp{\frac{\varv^2 t}{4 D}} \Bigg[ 2 z^{-1} \exp{-\frac{\epsilon \varv}{2D}} - \exp{\frac{\epsilon \varv}{2D}}\times\\ \nonumber
    &\erfc{\frac{\epsilon-\epsilon_i+t\varv}{\sqrt{4Dt}} } - \exp{\frac{\varv \epsilon_i}{D} -\frac{\epsilon \varv}{2D}}\times\\ \nonumber
   &\erfc{\frac{\epsilon_i-\epsilon+t\varv}{\sqrt{4Dt}}} \Bigg]\,. 
   \label{eq:partitionNonfixed}
\end{align}
To simplify the partition function, the energy-independent prefactors can be dropped because of the logarithmic derivative in Eq.\,\eqref{eq:Nformula}. We have used the complementary error function
\begin{align}
    \erfc{x} \equiv 1-\erf{x} \equiv \frac{2}{\sqrt{\pi}} \int_x^\infty e^{-t^2} \dd{t}\,,
\end{align}
with the normal error function \(\erf{x}\) that integrates from \(0\) up to \(x\) over the Gaussian distribution.

To obtain the occupation-number distribution \(n(\epsilon , t)\), the logarithmic derivative of \(\mathcal{Z}(\epsilon , t)\) must be evaluated. Therefore the energy derivative of the partition function will be calculated next. It can be shown that the derivatives with respect to \(\epsilon\) of the error functions in Eq.\,\eqref{eq:partitionNonfixed} equal zero,
\begin{align}
    \exp{\frac{\epsilon \varv}{2 D}} &\pdv{\epsilon} \erfc{\frac{\epsilon-\epsilon_i +t\varv}{\sqrt{4Dt}}} \notag \\
    +{}& \, \exp{\frac{\varv \epsilon_i}{D}-\frac{\epsilon \varv}{2D}} \pdv{\epsilon} \erfc{\frac{\epsilon_i-\epsilon+t\varv}{\sqrt{4Dt}}}  = 0\,.
\end{align}
This relation reduces the derivative of the partition function to the derivatives of the exponential functions, and the occupation-number distribution becomes 
\begin{align}
    n(\epsilon,t) ={}& \frac{1}{\exp{\frac{\epsilon-\mu}{T}} K(\epsilon,t)-1}
    \label{eq:SolutionNonfixedShortenedK}
    \end{align}
    with
\begin{align}
    K(\epsilon,t) \equiv \frac{2-\exp{\frac{\mu-\epsilon_i}{T}} \erfc{\frac{\epsilon_i-\epsilon+t\varv}{\sqrt{4Dt}}}}{\erfc{\frac{\epsilon-\epsilon_i+t\varv}{\sqrt{4Dt}}}}\,.
\end{align}
This result
already looks similar to the equilibrium Bose-Einstein distribution. 
A valid solution requires that Eq.\,\eqref{eq:SolutionNonfixedShortenedK} is equal to the initial distribution Eq.\,\eqref{eq:initialdistribution} for \(t \to 0\) and the Bose-Einstein distribution Eq.\,\eqref{Bose-Einstein} for \(t \to \infty\). This is the case if \(K(\epsilon,t)\) satisfies the following limits 
\begin{align}
    \lim \limits_{t \to 0} K(\epsilon,t) &\overset{!}{=}
    \begin{cases}
    1 &  \epsilon<\epsilon_i \\
    \infty &  \epsilon>\epsilon_i
    \end{cases} \\
    \lim \limits_{t \to \infty} K(\epsilon,t) &\overset{!}{=} C\,.
\end{align}
The limit for large times is left unspecified because the solution must have the structure of a Bose-Einstein distribution, but the chemical potential might change, which is denoted by a constant \(C\).
Using the limiting behaviour of the complementary error functions for \(t \to 0\)
\begin{align}
    \lim \limits_{t \to 0} \erfc{\frac{\epsilon-\epsilon_i+t\varv}{\sqrt{4Dt}}} &=
    \begin{cases}
    0 &  \epsilon>\epsilon_i \\
    2 &  \epsilon<\epsilon_i
    \end{cases} \\
    \lim \limits_{t \to 0} \erfc{\frac{\epsilon_i-\epsilon+t\varv}{\sqrt{4Dt}}} &=
    \begin{cases}
    0 &  \epsilon<\epsilon_i \\
    2 &  \epsilon>\epsilon_i\,,
    \end{cases} 
\end{align}
the correct limit for \(K(\epsilon,t)\) is obtained. For \(t \to \infty\) the correct limiting behaviour of the auxiliary function is also obtained,
\begin{align}
    \lim \limits_{t \to \infty} \erfc{\frac{\epsilon-\epsilon_i+t\varv}{\sqrt{4Dt}}} = 2 = \lim \limits_{t \to \infty} \erfc{\frac{\epsilon_i-\epsilon+t\varv}{\sqrt{4Dt}}}\,.
\end{align}
The latter equality holds for \(\varv<0\), which is fulfilled throughout this paper. Hence, the limits of the particle distribution 
\begin{align}
    \lim \limits_{t \to 0} n(\epsilon,t) &= n_{\mathrm{i}} (\epsilon , t) \\
    \lim \limits_{t \to \infty} n(\epsilon,t) &= \frac{1}{\exp{-\frac{\epsilon \varv}{D}} z^{-1} \underbrace{ \left( 1 - z \exp{\frac{\epsilon_i \varv}{D}}\right)}_{C} -1}
    \label{eq:stationarysolutionNonfixed}
\end{align}
obey the expected behaviour. In Eq.\,\eqref{eq:stationarysolutionNonfixed}, the constant \(C\) can be read off, which is the second difference to the initial distribution apart from the truncation: During the equilibration the thermal tail develops and approaches the Bose-Einstein distribution, whereas the singularity, in other words the chemical potential, moves to a new value \(\mu'\),
\begin{align}
    \mu' = \frac{D}{\varv} \ln{z^{-1} - \exp{\frac{\epsilon_i \varv}{D}}}\,.
\end{align}
The shift has no particular physical interpretation, it is a side effect of the choice of the free Green's function Eq.\,\eqref{eq:Greensnonfixed} without boundary conditions. The expression for \(\mu'\) shows that its value is not restricted to negative values, contradicting the requirement of negative chemical potentials in bosonic systems. 
Moreover, the shift causes a violation of particle-number conservation. Hence, this particular result does not provide a physically reasonable description. It is included here
to compare it with the one for proper boundary conditions at $\epsilon = \mu$.
\subsection{Fixed chemical potential}
\label{subsec:FixedchemPot}
As discussed in Sec.\,\ref{sec:solutionofnonlinearPDE}, another solution for the partition function Eq.\,\eqref{eq:newformulaforZ} is possible when we use boundary conditions at the singularity. Again, the initial particle distribution is a truncated Bose-Einstein distribution at temperature \(T =-\frac{D}{\varv}\), and the primitive \(A_{\mathrm{i}} (x)\) as well as \(F(x)\) do not change. The difference compared to Sec.\,\ref{subsec:VariableChemPot} arises in the formula for the partition function Eq.\,\eqref{eq:newformulaforZ} where the Green's function 
\(\Tilde{F}(\epsilon, x, t)\) is used and \(F(x)\) is evaluated at \(x+\mu\). The new Green's function maintains the singularity and thus, the chemical potential at a fixed value: In contrast to the previous approach, no unphysical shift in the chemical potential will occur.

As in the calculation for variable chemical potential, the integral has to be split into two sections separated by \(\epsilon_i - \mu\),
\begin{widetext}
\begin{align}
    \Tilde{\mathcal{Z}} (\epsilon,t) ={}& \int_0^\infty \Tilde{G} \left(\epsilon,x,t \right) F(x+\mu) \dd{x} = \int_0^\infty \Tilde{G} \left( \epsilon,x,t \right) \exp{\frac{x-\mu}{2 T}} \dd{x} \notag \\ 
    &- \int_0^{\epsilon_i-\mu} \Tilde{G} \left( \epsilon,x,t \right) \exp{-\frac{x+\mu}{2 T}} \dd{x} - \exp{-\frac{\epsilon_i}{T}} \int_{\epsilon_i-\mu}^\infty \Tilde{G} \left( \epsilon,x,t \right) \exp{\frac{x+\mu}{2 T}} \dd{x} \notag \\
    ={}& \sqrt{4Dt} \, \exp{\frac{\varv^2 t}{4D}-\frac{\mu}{2T}} \left( \exp{\frac{\epsilon-\mu}{2 T}} \Sigma_1 (\epsilon,t) - \exp{\frac{\mu-\epsilon}{2 T}} \Sigma_2 (\epsilon,t) \right)\,.
    \label{eq:Zfixedmu}
\end{align}
\end{widetext}
We have introduced two auxiliary functions \(\Sigma_1 (\epsilon,t)\) and \(\Sigma_2 (\epsilon ,t)\) which can be expressed in terms of complementary error functions and exponentials as
\begin{align}
    \Sigma_1 (\epsilon,t) \equiv{}& \erfc{\frac{2\mu - \epsilon_i -\epsilon +t \varv}{\sqrt{4 D t}}} \nonumber \\ 
    &- \exp{\frac{\mu-\epsilon_i}{T}} \erfc{\frac{\epsilon_i-\epsilon +t\varv}{\sqrt{4 D t}}} \nonumber\,,\\ 
    \end{align}
    and
    \begin{align}
    \Sigma_2 (\epsilon,t) \equiv{}& \erfc{\frac{\epsilon-\epsilon_i +t \varv}{\sqrt{4 \, D \, t}}} \notag \\ &- \exp{\frac{\mu-\epsilon_i}{T}} \erfc{\frac{\epsilon-2 \mu +\epsilon_i +t \varv}{\sqrt{4 D t}}}\,.
\end{align}
Due to the logarithmic derivative in Eq.\,\eqref{eq:Nformula}, we neglect energy-independent prefactors of the partition function. 

With this partition function and equation Eq.\,\eqref{eq:Nformula} for $\mathcal{Z} \rightarrow \Tilde{\mathcal{Z}}$, we calculate the occupation-number distribution. As in Sec.\,\ref{subsec:VariableChemPot}, the derivatives of the auxiliary functions with respect to \(\epsilon\) vanish
\begin{align}
    \exp{\frac{\epsilon-\mu}{2 T}} \pdv{\epsilon} \Sigma_1 (\epsilon,t) - \exp{\frac{\mu-\epsilon}{2 T}} \pdv{\epsilon} \Sigma_2 (\epsilon,t) =0\,.
\end{align}
Therefore, only the derivatives of the exponential functions are considered. The result
\begin{align}
    n(\epsilon,t) = \frac{1}{\exp{\frac{\epsilon-\mu}{T}} L(\epsilon,t)-1}
    \label{eq:particledistributionfixedmu}
\end{align}
is again similar to a Bose-Einstein distribution. In Eq.\,\eqref{eq:particledistributionfixedmu} the auxiliary function \(L(\epsilon,t)\) is used to absorb all terms that are responsible for the equilibration process
\begin{align}
    L(\epsilon,t) \equiv \frac{\Sigma_1 (\epsilon ,t)}{\Sigma_2 (\epsilon ,t)}\,.
\end{align}
To verify that the distribution function indeed describes the equilibration from an initial nonequilibrium distribution to a Bose-Einstein distribution, we analyze again the limiting behaviour for \(t \to 0\) and for \(t \to \infty\). The following limits of the sigma functions for negative \(\varv\) and \(\mu\) are used to determine the limiting behaviour of \(L(\epsilon,t)\)
\begin{align}
    \lim \limits_{t \to 0} \Sigma_1 (\epsilon,t) &= 2 - 2 \, \exp{\frac{\mu - \epsilon_i}{T}} \Theta (\epsilon - \epsilon_i) \\
    \lim \limits_{t \to 0} \Sigma_2 (\epsilon,t) &= 2 \, \Theta (\epsilon_i - \epsilon) \\
    \lim \limits_{t \to \infty} \Sigma_1 (\epsilon,t) &= 2 - 2 \, \exp{\frac{\mu - \epsilon_i}{T}} \\
    \lim \limits_{t \to \infty} \Sigma_2 (\epsilon,t) &= 2 - 2 \, \exp{\frac{\mu - \epsilon_i}{T}}\,.
\end{align}
Using these relations, the limits of \(L(\epsilon,t)\) can be determined
\begin{align}
    \lim \limits_{t \to 0} L(\epsilon,t) &= \begin{cases}
    1 &  \epsilon<\epsilon_i\\
    \infty &  \epsilon>\epsilon_i \end{cases} \\
    \lim \limits_{t \to \infty} L(\epsilon,t) &= 1
\end{align}
which show the correct behaviour. It can be concluded that for \(t \to 0\) the solution Eq.\,\eqref{eq:particledistributionfixedmu} equals the initial distribution Eq.\,\eqref{eq:initialdistribution} as expected. This is obvious for energies smaller than the truncation energy, whereas for higher energies the divergence of \(L(\epsilon,t)\) provides the vanishing distribution above \(\epsilon_i\). For the limit \(t \to \infty\) one arrives at an equilibrium Bose-Einstein distribution with temperature \(T\), which verifies the expectation.

When compared to Eq.\,\eqref{eq:stationarysolutionNonfixed}, the limit for \(t \to \infty\) equals one and thus, no shift in the chemical potential appears. This is in accordance with the approach that uses the modified Green's function Eq.\,\eqref{eq:newGreens}. However, we will show in Sec.\,\ref{sec:ParticleNumberConservation} that also this approach does not obey particle-number conservation. In the next section, both solutions will be analyzed and discussed.
\subsection{Analysis and discussion of both approaches}
\label{subsec:AnalysisandDiscussionofBothApproaches}
We now use the solutions from the previous calculations to analyze the equilibration behaviour of a cold bosonic gas in concrete examples, and discuss the differences and similarities between the two approaches without and with boundary conditions at the singularity.

For the analysis, a specific choice of parameters has to be made. To be able to compare directly with our earlier results in Ref.\,\cite{gw18a} with restricted initial conditions, we use the same parameters in this work. The transport coefficients are \(D = \SI{8000}{\pico \electronvolt^2 \second^{-1}}\) and \(\varv = \SI{-1000}{\pico \electronvolt \second^{-1}}\), resulting in an equilibrium temperature \(T = \SI{8}{\pico \electronvolt}\simeq 93\)\,nK. These values are motivated by experimental results for temperatures and time scales in ultracold $^{87}$Rb. At \(\epsilon_i = \SI{7}{\pico \electronvolt}\), the initial thermal distribution is truncated, and the chemical potential is chosen as \(\mu = \SI{-0.68}{\pico \electronvolt}\). 

With this set of parameters, the first solution Eq.\,\eqref{eq:SolutionNonfixedShortenedK} can be evaluated at different times \(t\).
\begin{figure}[t!]
    \centering
    \includegraphics[width = 0.5\textwidth]{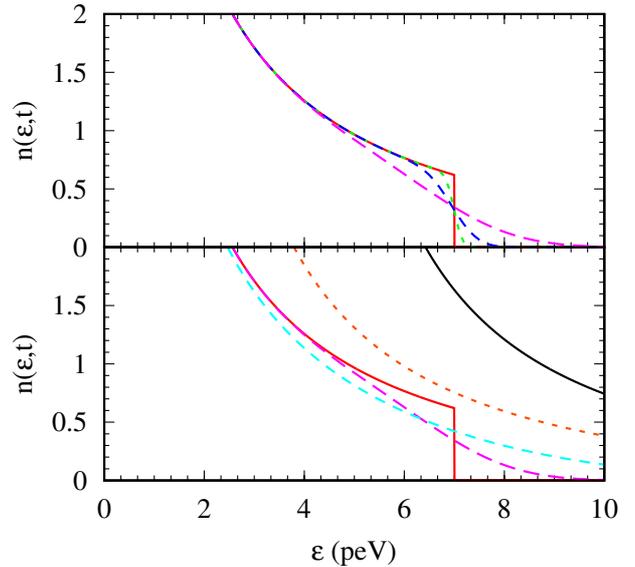}
    \caption{Time-dependent occupation-number distribution without boundary conditions describing the equilibration for variable \(\mu\) from an initial distribution at $t=0$
    (solid, red) to a Bose-Einstein distribution (solid, black). In the upper part $t=\,$\(\SI{1}{\micro \second}\) (short-dashed, green), \(\SI{10}{\micro \second}\) 
    (medium-dashed, blue) and \(\SI{100}{\micro\second}\) (long-dashed, purple) are displayed;   in the lower part $t=\,$\(\SI{100}{\micro\second}\) (long-dashed, purple), \(\SI{1}{\milli \second}\) (medium-dashed, turquoise) and \(\SI{10}{\milli\second}\) (short-dashed, orange) are shown together with the equilibrium distribution.}
    \label{fig:NonFixMuSmallTimes}
\end{figure}
From Fig.\,\ref{fig:NonFixMuSmallTimes} we conclude that the initial part of the equilibration (upper frame), where the discontinuity vanishes and the cutoff is smoothened, happens on the microsecond scale. At $t \simeq$ \(\SI{100}{\micro \second}\) a thermal tail above the cutoff energy has developed and a particle flow from energies below the cutoff to higher energies emerges. In comparison to the results in Ref.\,\cite{gw18a} where the initial distribution was confined to $\epsilon \ge 0$, one observes that on these short time scales both results are nearly identical. This implies that the equilibration behaviour at the cutoff \(\epsilon_i\) only depends on the local environment and not on the global distribution. For energies far below the truncation energy, no time evolution on short time scales occurs because here the particles are initially already Bose-Einstein-like distributed. 

The nearly symmetric equilibration below \(\SI{10}{\micro \second}\) around the cutoff energy supports the assumption that at the beginning, only local discontinuities are smoothed out. Fermions distributed according to a step function \cite{bgw19} show a similar equilibration behaviour. Moreover, at \(\epsilon = \epsilon_i\) they have a fixed point at half the particle density which does not hold for bosons. However, below \(t = \SI{100}{\micro \second}\) the particle density at the truncation does not change significantly even for bosons.

When reaching the millisecond scale, a global change of the distribution occurs. As discussed in Sec.\,\ref{subsec:VariableChemPot}, the equilibrium solution for variable chemical potential will be displaced compared to the initial distribution. This can be seen in the lower frame of Fig.\,\ref{fig:NonFixMuSmallTimes} where the distributions migrate horizontally toward the equilibrium solution. 
The new chemical potentials of these solutions become positive for $t\gtrsim$ \(\SI{10}{\milli \second}\), whereas the initial chemical potential is negative. This indicates that the total number of thermal bosons increases through a shift toward higher energies and becomes infinite at \(\mu = 0\).
Consequently, these solutions do not fulfill particle-number conservation as they are supposed to do.  Despite this deficiency, the distributions approach a non-zero equilibrium solution for large times. Hence, the model does describe the transition from a non-equilibrium initial condition to an equilibrium solution.
\begin{figure}[t!]
    \centering
    \includegraphics[width = 0.50\textwidth]{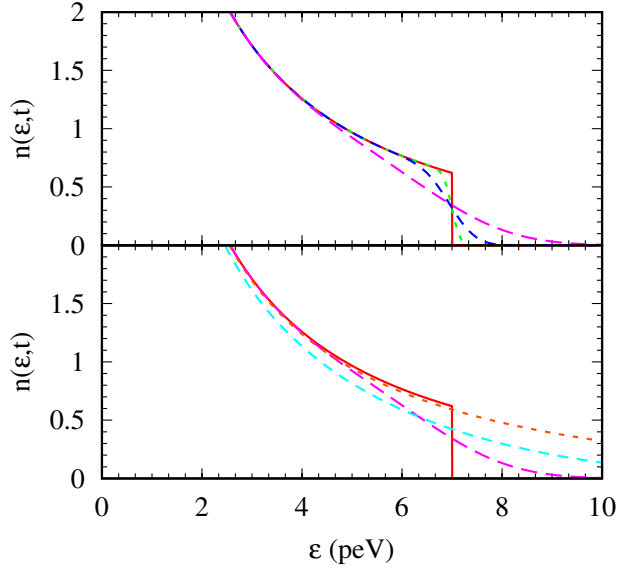}
    \caption{Time-dependent occupation-number distribution including boundary conditions for fixed \(\mu\) describing the equilibration from an initial distribution at 
 $t=0$   (solid, red) to a Bose-Einstein distribution. In the upper part $t\,=\,$\(\SI{1}{\micro \second}\) (short-dashed, green), \(\SI{10}{\micro \second}\) (medium-dashed, blue) and \(\SI{100}{\micro\second}\) (long-dashed, purple) are displayed;   in the lower part $t\,=\,$\(\SI{100}{\micro\second}\) (long-dashed, purple), \(\SI{1}{\milli \second}\) (medium-dashed, turquoise) and \(\SI{10}{\milli\second}\) (short-dashed, orange) are shown.}
    \label{fig:FixMuSmallTimes}
\end{figure}

Using the same set of parameters, we now investigate the solution Eq.\,\eqref{eq:particledistributionfixedmu} with fixed chemical potential in Fig.\,\ref{fig:FixMuSmallTimes}. On the microsecond scale (upper frame), there is no visual difference between these distributions and those in Fig.\,\ref{fig:NonFixMuSmallTimes}. The equilibration also starts with a local disappearance of the discontinuity in a nearly symmetric way around the truncation energy. Hence, the choice of fixed or variable chemical potential has no significant influence on the equilibration on these time scales. 

In contrast to the small time scales, for large times (lower frame of Fig.\,\ref{fig:FixMuSmallTimes}) an important difference to the results with variable chemical potential is observed. The final distribution has the same singularity as the initial one, hence no horizontal movement toward higher energies occurs. 
Instead the thermal tail grows and results in the correct Bose-Einstein distribution with negative chemical potential $\mu<0$. 

 To quantify the time scale that separates the short- and the long-time regions, a separation time $t_\text{s}$ should be found. Besides the vague characterization that the initial part occurs on the microsecond and the final one on the millisecond scale, a more precise separation scale can be derived. For short times the particle distribution has two inflection points, one nearly at the truncation energy and another one at a slightly smaller energy. These points disappear at a particular time \(t_{\mathrm{s}}\) because the equilibrium solution is convex for positive energies. For both solutions and for the present set of parameters, we find \(t_{\mathrm{s}} \approx \SI{100}{\micro \second}\), supporting the observation that on short time scales the solutions behave similar. 
 For $t > t_\text{s}$,
 the inflection points are smoothened, the thermal tail rises and the equilibrium solution is approached. So far, the initial temperature $T$ has not yet changed.

\section{Equilibration for arbitrary initial temperatures}
\label{sec:EquilforArbitraryInitialTemp}
We now investigate the equilibration process for an arbitrary initial temperature, which may change during the thermalization.  This enables the investigation of an actual cooling process from an initial temperature \(T_{\mathrm{i}}\) to a final temperature \(T_{\mathrm{f}} = -\frac{D}{\varv}\), determined by the transport coefficients. Both temperatures are parameters:
If the model is used to describe a cooling process as in actual cold-atom experiments, the temperatures have to be chosen according to the experimental properties.
\subsection{Calculation of the partition function}
\label{subsec:CalcofPartFunction}
To account schematically for evaporative cooling, we use again the truncated Bose-Einstein distribution as initial distribution, but now the initial temperature $T \equiv $ \(T_{\mathrm{i}}\) is not determined by the transport coefficients anymore.
For the calculation of the partition function \(\mathcal{Z}(\epsilon,t)\) Eq.\,\eqref{eq:newformulaforZ} is used, thus including the boundary condition at the singularity: The chemical potential is treated as a parameter that does not change, it determines the singularity in the particle distribution for all times. 

With this initial condition, the primitive \(A_{\mathrm{i}}(x)\) stays the same as in Eq.\,\eqref{eq:antiderivative} and \(F(x+\mu)\) can directly be evaluated as
\begin{align}
    F(x&+\mu) = \notag \\
&= \begin{cases}
    \exp{ \frac{x+\mu}{2 T_{\mathrm{f}}} + \frac{1}{T_{\mathrm{f}}} \int_0^{x+\mu} n_i(y) \dd{y} } &  x<\epsilon_i-\mu \\
    \exp{\frac{x+\mu}{2 T_{\mathrm{f}}} + \frac{1}{T_{\mathrm{f}}} \int_0^{\epsilon_i} n_i(y) \dd{y} } &  x>\epsilon_i-\mu
    \end{cases} \notag\\
    &= \begin{cases}
    \exp{\frac{x-\mu}{2 T_{\mathrm{f}}}} \left( 1-\exp{-\frac{x}{T_{\mathrm{i}}}} \right)^{\frac{T_{\mathrm{i}}}{T_{\mathrm{f}}}} &  x<\epsilon_i-\mu \\
    \exp{\frac{x-\mu}{2 T_{\mathrm{f}}}} \left( 1-\exp{\frac{\mu-\epsilon_i}{T_{\mathrm{i}}}} \right)^{\frac{T_{\mathrm{i}}}{T_{\mathrm{f}}}} &  x>\epsilon_i-\mu\,.
    \end{cases}
\end{align}
In contrast to Eq.\,\eqref{eq:F(x)}, the prefactor \(T_{\mathrm{i}}\) of the primitive does not vanish and yields the exponent occuring in \(F(x+\mu)\). This quantity is a real number, because the exponential function within the brackets is smaller than one for both energy regions. As in Sec.\,\ref{sec:SolutionforTruncEquilDistr}, the two cases in \(F(x+\mu)\) force the separation of the partition function in two parts denoted by \(\mathcal{Z}_1 (\epsilon ,t)\) and \(\mathcal{Z}_2 (\epsilon ,t)\),
\begin{align}
    \Tilde{\mathcal{Z}} (\epsilon,t) ={}& \underbrace{ \int_0^{\epsilon_i-\mu} \Tilde{G} \left( \epsilon,x,t \right) F(x+\mu) \dd{x} }_{\mathcal{Z}_1 (\epsilon ,t)} \notag \\
    &+ \underbrace{ \int_{\epsilon_i-\mu}^\infty \Tilde{G} \left( \epsilon,x,t \right) F(x+\mu) \dd{x} }_{\mathcal{Z}_2 (\epsilon,t)}
\end{align}
where \(\mathcal{Z}_1 (\epsilon,t)\) can be expressed as
\begin{align}
    \mathcal{Z}_1 (\epsilon,t) ={}& \int_0^{\epsilon_i-\mu} \Tilde{G} \left( \epsilon,x,t \right) \exp{\frac{x-\mu}{2 T_{\mathrm{f}}}} \times \notag \\
    &\left( 1-\exp{-\frac{x}{T_{\mathrm{i}}}} \right)^{\frac{T_{\mathrm{i}}}{T_{\mathrm{f}}}} \dd{x} \notag \\
    ={}& \sum_{k=0}^\infty \binom{\frac{T_{\mathrm{i}}}{T_{\mathrm{f}}}}{k} \left( -1\right)^k \exp{-\frac{\mu}{2 T_{\mathrm{f}}}} \times \notag \\ 
    &\underbrace{ \int_0^{\epsilon_i-\mu} \Tilde{G} \left( \epsilon,x,t \right) \expo{\alpha_k x} \dd{x}}_{\mathcal{Z}_1^{k}(\epsilon,t)}\,.
    \label{eq:Z1calc}
\end{align}
Here we have used the generalized binomial theorem that can be derived from a Taylor expansion around \(x = 0\), 
\begin{align}
    (1+x)^s = \sum_{k=0}^\infty \binom{s}{k} \, x^k\,.
    \label{eq:Binomial}
\end{align}
It holds for real numbers \( s \in \mathbb{R} \) and converges absolutely if \( \abs{x}<1 \). In Eq.\,\eqref{eq:Binomial} the generalized binomial coefficient
\begin{align}
    \binom{s}{k} \equiv \frac{s \left( s-1\right) \dots \left( s-k+1 \right) }{k!} 
\end{align}
is introduced. Because the absolute value of \(\exp{-\frac{x}{T_{\mathrm{i}}}}\) is smaller than one, the theorem can be applied. To simplify Eq.\,\eqref{eq:Z1calc}, the auxiliary function \(\alpha_k \equiv \frac{1}{2 T_{\mathrm{f}}} - \frac{k}{T_{\mathrm{i}}} \) is used. In the next step the coefficients \(\mathcal{Z}_1^k (\epsilon ,t) \) that were defined in Eq.\,\eqref{eq:Z1calc} are calculated as
\begin{align}
    \mathcal{Z}_1^{k} ={}& \int_0^{\epsilon_i-\mu} \Bigg[ \exp{-\frac{(\epsilon-\mu-x)^2}{4 D t}} \notag \\
    &\qquad \qquad \quad - \exp{-\frac{(\epsilon-\mu+x)^2}{4 D t}} \Bigg] \expo{\alpha_k x} \dd{x} \notag \\
    ={}& \sqrt{\pi D t} \, \expo{\alpha_k^2 D t} \left[ \expo{\alpha_k (\epsilon-\mu)} \Lambda_1^{k} (\epsilon,t) - \expo{\alpha_k (\mu -\epsilon)} \Lambda_2^{k} (\epsilon,t) \right]\,.
    \label{Z1alpha}
\end{align}
The two auxiliary functions \(\Lambda_1^{k} (\epsilon,t)\) and \( \Lambda_2^{k} (\epsilon,t) \) are defined with a superscript \(k\), expressing the \(k\) dependance of both auxiliary functions
\begin{align}
    \Lambda_1^{k} (\epsilon,t) \equiv{}& \erf{\frac{\epsilon - \mu +2 D t \alpha_k}{\sqrt{4 D t}}} - \erf{\frac{\epsilon - \epsilon_i + 2 D t \alpha_k}{\sqrt{4 D t}}} \notag \\
    \Lambda_2^{k} (\epsilon,t) \equiv{}& \erf{\frac{\mu-\epsilon+ 2 D t \alpha_k}{\sqrt{4 D t}}} \notag \\ 
    &\qquad - \erf{\frac{2 \mu - \epsilon - \epsilon_i + 2 D t \alpha_k}{\sqrt{4 D t}}}\,.
\end{align}
The second summand \(\mathcal{Z}_2 (\epsilon ,t)\) can be calculated without the generalized binomial theorem, due to the missing \(x\) dependence in the second factor in \(F(x+ \mu )\)
\begin{align}
    \mathcal{Z}_2 ={}& \left( 1- \exp{\frac{\mu-\epsilon_i}{T_{\mathrm{i}}}} \right)^{\frac{T_{\mathrm{i}}}{T_{\mathrm{f}}}} \times \notag \\
    &\qquad \int_{\epsilon_i-\mu}^\infty \Tilde{G} \left( \epsilon,x,t \right) \exp{\frac{x-\mu}{2 T_{\mathrm{f}}}}  \dd{x} \notag \\
    ={}& \left( 1- \exp{\frac{\mu-\epsilon_i}{T_{\mathrm{i}}}} \right)^{\frac{T_{\mathrm{i}}}{T_{\mathrm{f}}}} \sqrt{\pi D t} \, \exp{\frac{D t}{4 T_{\mathrm{f}}^2} - \frac{\mu}{2 T_{\mathrm{f}}}} \times \notag \\
    & \Bigg[ \exp{\frac{\epsilon-\mu}{2 T_{\mathrm{f}}}} \Lambda_3 (\epsilon,t) - \exp{\frac{\mu-\epsilon}{2 T_{\mathrm{f}}}} \Lambda_4 (\epsilon,t) \Bigg]\,.
\end{align}
Two additional auxiliary functions \(\Lambda_3 ( \epsilon,t)\) and \(\Lambda_4 ( \epsilon,t)\) are defined
\begin{align}
    \Lambda_3 ( \epsilon,t) &\equiv \erfc{\frac{\epsilon_i - \epsilon + t \varv }{\sqrt{4 D t}}} \\
    \Lambda_4 ( \epsilon,t) &\equiv \erfc{\frac{\epsilon - 2 \mu + \epsilon_i + t \varv}{\sqrt{4 D t}}}
\end{align}
to abbreviate the expression and absorb the complementary error functions.

With \(\mathcal{Z}_1^k (\epsilon ,t)\), \(\mathcal{Z}_2 (\epsilon ,t)\) and the four auxiliary functions, the total partition function   $ \Tilde{\mathcal{Z}}  (\epsilon ,t))$ becomes
\begin{align}
    \Tilde{\mathcal{Z}}  ={}& \sqrt{4 D t} \, \exp{-\frac{\mu}{2 T_{\mathrm{f}}}} \sum_{k=0}^{\infty} \binom{\frac{T_{\mathrm{i}}}{T_{\mathrm{f}}}}{k} \left( -1 \right)^k \times  \notag \\
    &\Bigg( \expo{\alpha_k^2 D t} \left[ \expo{\alpha_k (\epsilon - \mu)} \Lambda_1^k (\epsilon,t) - \expo{\alpha_k (\mu - \epsilon)} \Lambda_2^k (\epsilon, t) \right] \notag \\
    &+ \exp{\frac{(\mu - \epsilon_i)k}{T_{\mathrm{i}}}}  \exp{\frac{D t}{4 T_{\mathrm{f}}^2}} \times \notag \\
    &\Bigg[ \exp{\frac{\epsilon-\mu}{2 T_{\mathrm{f}}}} \Lambda_3 (\epsilon,t) - \exp{\frac{\mu - \epsilon}{2 T_{\mathrm{f}}}} \Lambda_4 (\epsilon,t) \Bigg] \Bigg)\,.
    \label{eq:Zarbtemp}
\end{align}
In this result \(\mathcal{Z}_2 (\epsilon ,t)\) is also represented as an infinite series. Like in the previous partition functions some negligible, energy independent prefactors appear. If the fraction between the initial and final temperature is integer, the infinite sum terminates after a finite number of summations, but in realistic physical situations the infinite sum has to be evaluated. For \(T_{\mathrm{f}} = T_{\mathrm{i}}\) only two summands are left, and the partition function Eq.\,\eqref{eq:Zfixedmu} is recovered. Hence, Eq.\,\eqref{eq:Zarbtemp} is the generalized partition function for arbitrary temperatures and fixed chemical potential. 
\subsection{Occupation-number distribution and limiting behaviour}
\label{subsec:ParticleDistransLimitBehaviour}
With the above result for the partition function, we proceed to calculate the occupation-number distribution \(n(\epsilon,t)\).
At first the derivatives of the complementary error functions are computed. It is again expected that the contribution will vanish and only the derivatives of the exponential functions must be taken into account
\begin{align}
    2 \, &\exp{-\frac{(\epsilon-\mu)^2}{4 D t}} = \notag \\
    &\expo{\alpha_k^2 D t} \left[ \expo{\alpha_k (\epsilon - \mu)} \, \pdv{\epsilon} \Lambda_1^k (\epsilon,t) - \expo{\alpha_k (\mu - \epsilon)} \, \pdv{\epsilon}\Lambda_2^k (\epsilon, t) \right] \notag \\
    &+ \exp{\frac{(\mu - \epsilon_i)k}{T_{\mathrm{i}}}}  \exp{\frac{D t}{4 T_{\mathrm{f}}^2}} \times \notag \\ 
    &\left[ \exp{\frac{\epsilon-\mu}{2 T_{\mathrm{f}}}} \pdv{\epsilon}\Lambda_3 (\epsilon,t) - \exp{\frac{\mu - \epsilon}{2 T_{\mathrm{f}}}} \pdv{\epsilon}\Lambda_4 (\epsilon,t) \right]\,.
\end{align}
The resulting term after the derivation of all four auxiliary functions is non-vanishing, but independent from the summation index \(k\). Therefore, it can be written in front of the summation and only the binomial coefficients and the powers of \(-1\) remain. In general, the binomial theorem only holds for \(\abs{x} < 1\) and hence, is not applicable to the result where \(x = -1\). However, for positive exponents \(s \in \mathbb{R}_{+}\) the statement of the generalized binomial theorem can be extended to arguments \(\abs{x} \le 1 \). Because the temperature fraction is positive, the theorem can be applied and finally holds the expected vanishing contribution
\begin{align}
    \sum_{k=0}^\infty \binom{\frac{T_{\mathrm{i}}}{T_{\mathrm{f}}}}{k} \left( -1 \right)^k = (1-1)^\frac{T_{\mathrm{i}}}{T_{\mathrm{f}}} = 0\,.
\end{align}
Therefore, the derivative of $\Tilde{\mathcal{Z}} (\epsilon ,t)$) can be calculated by deriving the exponential functions
\begin{align}
   \pdv{\epsilon} \Tilde{\mathcal{Z} }={}& \sqrt{4 D t} \, \exp{-\frac{\mu}{2 T_{\mathrm{f}}}} \sum_{k=0}^{\infty} \nonumber
   \binom{\frac{T_{\mathrm{i}}}{T_{\mathrm{f}}}}{k} \left( -1 \right)^k  \times\\ \nonumber
   &\Bigg( \alpha_k \expo{\alpha_k^2 D t} \left[ \expo{\alpha_k (\epsilon - \mu)} \Lambda_1^k (\epsilon,t)\nonumber
    + \expo{\alpha_k (\mu - \epsilon)} \Lambda_2^k (\epsilon, t) \right] \nonumber \\
    &+ \exp{\frac{(\mu - \epsilon_i)k}{T_{\mathrm{i}}} + \frac{D t}{4 T_{\mathrm{f}}^2}} \frac{1}{2T_{\mathrm{f}}} \times\\ \nonumber
    &\left[ \exp{\frac{\epsilon-\mu}{2 T_{\mathrm{f}}}} \Lambda_3 (\epsilon,t) + \exp{\frac{\mu - \epsilon}{2 T_{\mathrm{f}}}} \Lambda_4 (\epsilon,t) \right] \Bigg)\,. \nonumber
    \label{eq:derivativeofZnewTemp}
\end{align}
With Eq.\,\eqref{eq:Nformula}, the full solution for an arbitrary initial temperature can now be deduced.

Besides the fact that the solution yields the correct result for \(T_{\mathrm{i}} = T_{\mathrm{f}}\), the limiting behaviour for \(t=0\) and for \(t \to \infty\) must be calculated in order to check the validity of the solution. At first the behaviour for \(t \to 0\) is investigated. This results in
\begin{align}
    \lim \limits_{t \to 0} \Lambda_1^k (\epsilon,t) &= 2 \, \Theta (\epsilon_i - \epsilon) \\
    \lim \limits_{t \to 0} \Lambda_2^k (\epsilon,t) &= 0 \\
    \lim \limits_{t \to 0} \Lambda_3 (\epsilon,t) &= 2 \, \Theta (\epsilon - \epsilon_i) \\
    \lim \limits_{t \to 0} \Lambda_4 (\epsilon,t) &= 0
\end{align}
for the auxiliary functions. With these values the limit of $\Tilde{ \mathcal{Z} }(\epsilon ,t) $ holds
\begin{align}
    \lim \limits_{t \to 0} \Tilde{\mathcal{Z}} (\epsilon,t) ={}& 2\, \exp{\frac{\epsilon - 2 \mu}{2 T_{\mathrm{f}}}} \times \notag \\
&\begin{cases} \left( 1 - \exp{\frac{\mu - \epsilon}{T_{\mathrm{i}}}} \right)^{\frac{T_{\mathrm{i}}}{T_{\mathrm{f}}}} &  \epsilon < \epsilon_i \\
    \left( 1 - \exp{\frac{\mu - \epsilon_i}{T_{\mathrm{i}}}} \right)^{\frac{T_{\mathrm{i}}}{T_{\mathrm{f}}}} &  \epsilon > \epsilon_i \,,\end{cases} 
\end{align}
where energy independent prefactors are neglected. In the expression the generalized binomial theorem is applied to return to a power expression without infinite summation.

Apart from the limit of the partition function itself, also the limit of the derivative of the partition function is necessary to obtain the initial distribution. It is sufficient to take the derivative of \( \lim \limits_{t \to 0} \Tilde{\mathcal{Z}} (\epsilon ,t) \) with respect to \( \epsilon \)
\begin{align}
    \lim \limits_{t \to 0} &\pdv{\epsilon} \Tilde{\mathcal{Z} }(\epsilon,t) =  \frac{1}{T_{\mathrm{f}}} \, 			\exp{\frac{\epsilon - 2 \mu}{2 T_{\mathrm{f}}}} \times \notag \\
&\begin{cases} \left( 1 - \exp{\frac{\mu - \epsilon}{T_{\mathrm{i}}}} \right)^{\frac{T_{\mathrm{i}}}{T_{\mathrm{f}}}} \frac{ 1 + \exp{\frac{\mu - \epsilon}{T_{\mathrm{i}}}}}{1 - \exp{\frac{\mu - \epsilon}{T_{\mathrm{i}}}}} &  \epsilon < \epsilon_i \\ \left( 1 - \exp{\frac{\mu-\epsilon_i}{T_{\mathrm{i}}}} \right)^{\frac{T_{\mathrm{i}}}{T_{\mathrm{f}}}} &  \epsilon > \epsilon_i\,. \end{cases}
\end{align}
With these results it can be shown that \(n(\epsilon,0)\) equals the initial distribution Eq.\,\eqref{eq:initialdistribution}.

For the calculation of the \(t \to \infty\) limit, the limits of the auxiliary functions are determined
\begin{align}
    \lim \limits_{t \to \infty} \Lambda_1^k (\epsilon,t) &= 0 \\ 
    \lim \limits_{t \to \infty} \Lambda_2^k (\epsilon,t) &= 0 \\
    \lim \limits_{t \to \infty} \Lambda_3 (\epsilon,t) &= 2 \\
    \lim \limits_{t \to \infty} \Lambda_4 (\epsilon,t) &= 2\,.
\end{align}
These limits are yet not sufficient to calculate the limit of the partition function. In front of every auxiliary function an exponential function occurs that diverges for large times. To investigate the peculiarities of the limit, it is useful to determine the limits of \(\mathcal{Z}_1 (\epsilon ,t)\) and \(\mathcal{Z}_2 (\epsilon ,t)\) separately.

For \(\Lambda_3 (\epsilon,t)\) and \(\Lambda_4 (\epsilon,t) \) the non-zero finite limits do not oppress the time-dependent exponential function. This means that \(\mathcal{Z}_2 (\epsilon ,t)\) does not converge for \(t \to \infty\) what is yet no problem because it will be later oppressed by the nominator in Eq.\,\eqref{eq:Nformula}.

When investigating the limit of \(\mathcal{Z}_1 (\epsilon ,t )\) it can be proven that the auxiliary functions decrease faster than the exponential prefactors increase and thereby vanish
\begin{align}
    \lim \limits_{t \to \infty} \expo{\alpha_k^2 D t} \Lambda_1^k (\epsilon ,t) &= 0 \\
    \lim \limits_{t \to \infty} \expo{\alpha_k^2 D t} \Lambda_2^k (\epsilon ,t) &= 0\,.
\end{align}
However, this is not sufficient since there is still a global prefactor \(\sqrt{4Dt}\) in front of the partition function. But when evaluating the limit of the logarithmic derivative, this prefactor drops out. As a consequence, all terms with \(\Lambda_1^k (\epsilon ,t)\) and \(\Lambda_2^k (\epsilon ,t)\) vanish and likewise \(\mathcal{Z}_1 (\epsilon ,t)\). Just the logarithmic derivative of \(\mathcal{Z}_2 (\epsilon ,t)\) is of importance when taking the limit,
\begin{align}
    \lim \limits_{t \to \infty} \pdv{\epsilon} \ln{\Tilde{\mathcal{Z}}} ={}& \lim \limits_{t \to \infty} \pdv{\epsilon}\ln{\mathcal{Z}_2} = \notag \\
={}&\frac{1}{2 T_{\mathrm{f}}} \, \frac{\exp{\frac{\epsilon-\mu}{2 T_{\mathrm{f}}}} + \exp{\frac{\mu - \epsilon}{2 T_{\mathrm{f}}}}}{\exp{\frac{\epsilon - \mu }{2 T_{\mathrm{f}}}} - \exp{\frac{\mu - \epsilon}{2 T_{\mathrm{f}}}}}\,.
\end{align}
With this result, the final distribution of \(n(\epsilon,t)\) can be determined using Eq.\,\eqref{eq:Nformula}
\begin{align}
    \lim \limits_{t \to \infty} n(\epsilon ,t) = \frac{1}{\exp{\frac{\epsilon- \mu}{T_{\mathrm{f}}}} - 1}\,.
\end{align}
Hence, a Bose-Einstein distribution with equilibrium temperature \(T_{\mathrm{f}}\) is obtained, and it is now possible to describe a cooling process from an arbitrary initial temperature to the equilibrium temperature.
\subsection{Discussion of the solutions}
\label{subsec:DiscussionoftheSol}
\begin{figure}[b!]
    \centering
    \includegraphics[width = 0.50\textwidth]{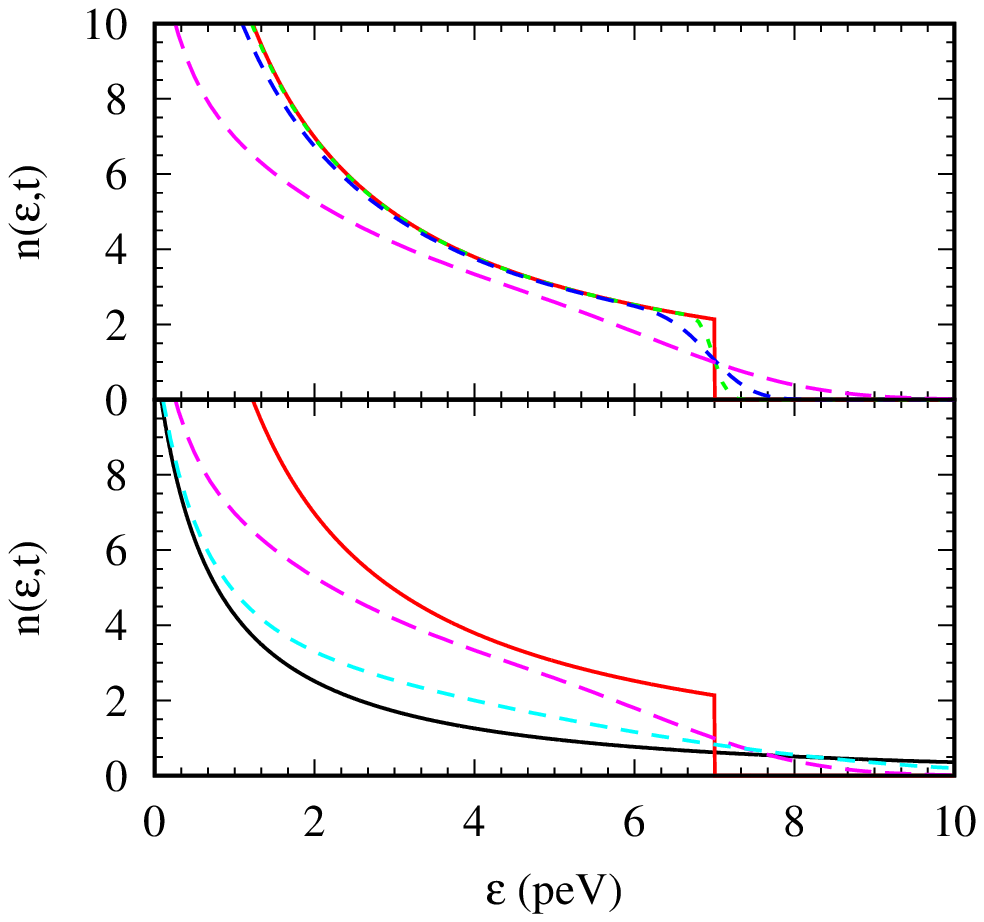}
    \caption{Time-dependent occupation-number distribution for fixed \(\mu\) describing the equilibration from an initial distribution at 
    $t=0$ and \(T=\SI{20}{\pico \electronvolt}\) (solid, red) to a Bose-Einstein distribution at \(T=\SI{8}{\pico \electronvolt}\) (solid, black). In the upper part $t=\,$\(\SI{1}{\micro \second}\) (short-dashed, green), \(\SI{10}{\micro \second}\) (medium-dashed, blue) and \(\SI{100}{\micro\second}\) (long-dashed, purple) are displayed;   in the lower part $t=\,$\(\SI{100}{\micro\second}\) (long-dashed, purple) and \(\SI{500}{\micro \second}\) (medium-dashed, turquoise) are shown together with the equilibrium distribution.}
    \label{fig:Temp20SmallTimes}
\end{figure}
We now apply the solutions with the same parameter set as in the previous section, but for arbitrary initial temperature $T_\text{i}$.
The temperature is chosen such that no integer temperature fraction causes a finite series. We start with an initial temperature 
\(T = \SI{20}{\pico \electronvolt}\) as shown in
 Fig.\,\ref{fig:Temp20SmallTimes}. The major difference to earlier equilibration plots is the change in the infrared, or low-energy regime. Already at \(\SI{10}{\micro \second}\) the distribution can be clearly distinguished from the initial distribution, whereas in Fig.\,\ref{fig:FixMuSmallTimes} the distribution still overlaps at these energies. However, on shorter time scales below \(\SI{10}{\micro \second}\) the equilibration resembles the equilibration in Fig.\,\ref{fig:NonFixMuSmallTimes} and \ref{fig:FixMuSmallTimes}.
 
The region around the truncation energy shows a similar equilibration behaviour compared to Sec.\,\ref{subsec:AnalysisandDiscussionofBothApproaches}. For short times, the same symmetric equilibration appears with an almost constant particle density at the truncation energy. An explanation for this similarity is again the locality of the equilibration such that both cases are similar up to a scaling factor.

Besides the equilibration around the discontinuity, the distribution has to move into a Bose-Einstein distribution at lower temperature. On the \(\SI{100}{\micro \second}\) scale this global redistribution takes place and the thermal tail develops. The distribution first approaches the equilibrium distribution at low energies and around the truncation, whereas the redistribution takes longer in the energy region between \(\SI{1}{\pico \electronvolt}\) and \(\SI{6}{\pico \electronvolt}\). However, when reaching the \(\si{\milli \second}\) scale, an equilibrium Bose-Einstein distribution at \(T_{\mathrm{f}} = \SI{8}{\pico \electronvolt}\) is approached.

As in Sec.\,\ref{subsec:AnalysisandDiscussionofBothApproaches}, the time where the inflection points vanish can be determined as well. This happens at \(t_{\mathrm{i}} \approx \SI{160}{\micro \second}\), which is somewhat later than in case of the solution for fixed temperature due to the early change at low energies. Again, the value can be used to divide the equilibration into two time regimes. First the discontinuity is smoothened and eventually vanishes; in the second part, the equilibrium solution is approached on every energy scale. In the latter part also the density distribution at the truncation energy changes and shrinks to its final value. This effect seems to be more characteristic to separate both parts, because already at times below \(t_{\mathrm{s}}\) a significant difference to the initial distribution becomes apparent.

\section{Particle-number conservation}
\label{sec:ParticleNumberConservation}
So far, all our model calculations violate particle-number conservation. Different from relativistic systems where particles can be created from the available energy, the particle number in cold bosonic gases must, of course, be conserved. The conservation law can be introduced by shifting the chemical potential of the distribution until the correct particle number is reached. In the previous sections, we have treated the chemical potential as a parameter, resulting in the violation of particle-number conservation. The main remaining task is therefore the numerical calculation of a time-dependent chemical potential \(\mu (t)\) in accordance with particle-number conservation.
\subsection{Calculation of the particle number}
\label{subsec:CalcoftheParticleNumber}
To include the particles in the ground state, the total particle number \(N\) has to be split up into two time-dependent contributions
\begin{align}
    N = N_{\mathrm{c}}(t) + N_{\mathrm{th}}(t)\,.
    \label{eq:particleNumberconservation}
\end{align}
The number of particles in the ground state is denoted by \(N_{\mathrm{c}} (t)\) and the number of particles in the thermal cloud by \(N_{\mathrm{th}} (t)\). For the purpose of the present investigation, we take the value of the distribution function Eq.\,\eqref{eq:Nformula} at \(\epsilon=0\)\, to represent the first contribution
\begin{align}
    N_{\mathrm{c}} (t) = n(\epsilon=0,t)\,.
\end{align}
Due to the negative initial value of the chemical potential, $\mu$ must be negative at all times. If it had become exactly zero, the number of particles would have diverged and therefore particle-number conservation been violated. Only in the thermodynamical limit below the critical temperature for condensation, \(\mu = 0\) is approached.

The number of particles in the thermal cloud is defined by the integral over the full distribution weighted with the density of states \(g(\epsilon)\)
\begin{align}
    N_{\mathrm{th}}(t) = \int_0^\infty g ( \epsilon) \, n ( \epsilon,t ) \dd{\epsilon}\,.
    \label{eq:thermalBosons}
\end{align}
It is assumed that the density of states obeys the power law
\begin{align}
    g(\epsilon) = g_0 \, \epsilon^k
\end{align}
where \(k \in \mathbb{R}\) and \( g_0 \) is an energy-independent constant. For an ideal uniform Bose gas without external potential the exponent \(k\) equals \(\frac{1}{2}\) and
\begin{align}
    g_0 = \frac{V}{4 \pi^2} \left(\frac{2 m}{\hbar^2}\right)^{\frac{3}{2}}\,.
\end{align}
It can be derived through the substitution of a summation over the quantum numbers of the associated states with an energy integration \cite{BookPitaevskii}.

The total number of particles at \(t=0\) of the truncated initial distribution can then be calculated as
\begin{align}
    N = \frac{1}{z^{-1}-1} + g_0 \int_0^{\epsilon_i} \frac{\sqrt{\epsilon}}{\exp{\frac{\epsilon-\mu}{T_{\mathrm{i}}}}-1} \dd{\epsilon}\,.
\end{align}
With the total particle number \(N\) and Eq.\,\eqref{eq:particleNumberconservation}, an implicit equation for \(\mu(t)\) is obtained. This equation can be evaluated numerically at every time \(t\) and thus, a time-dependent chemical potential is derived.

However, by imposing a time dependence in the chemical potential, the distributions functions calculated in Secs.\,\ref{sec:SolutionforTruncEquilDistr} and \ref{sec:EquilforArbitraryInitialTemp} become approximate solutions of Eq.\,\eqref{bose}: When taking the time derivative, new terms will arise that do not vanish. A fully self-consistent approach to calculate the time-dependent chemical potential and a detailed quantum treatment of the condensate fraction is beyond the scope of this work.
\subsection{Evaluation for \(T_{\mathrm{i}} = T_{\mathrm{f}}\)}\label{subsec:EvaluationforTi=Tf}
We now calculate the time-dependent solutions of the NBDE using the formulas from the previous section, first with equal initial and final temperatures. The parameters are the same as in Sec.\,\ref{subsec:AnalysisandDiscussionofBothApproaches}, but supplemented by \(g_0 = \SI{100}{\pico \electronvolt^{-\frac{3}{2}}}\), and \(k = \frac{1}{2}\) for the density of states in a three-dimensional isotropic Bose gas. 
Here we have estimated \(g_0\) to describe systems comparable to a dilute $^{87}$Rb vapor \cite{an95}, but in our case with only several thousand atoms.
\begin{figure}[bt]
    \centering
    \includegraphics[width = 0.50\textwidth]{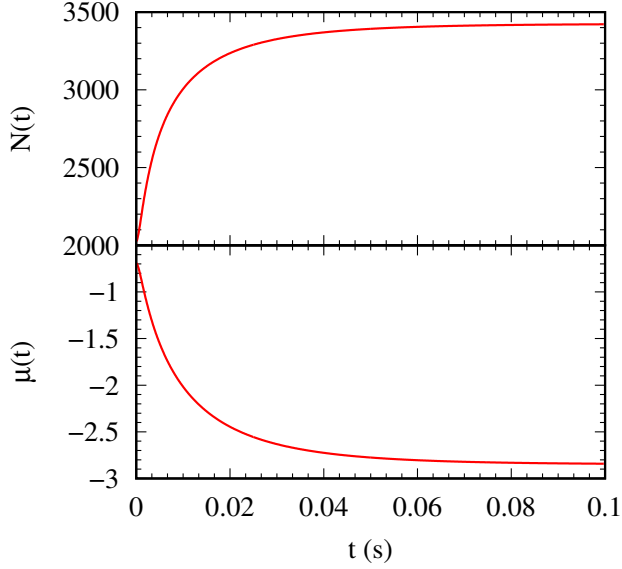}
    \caption{Particle number for fixed chemical potential \(\mu = \SI{-0.68}{\pico \electronvolt}\) and chemical potential for conserved particle number.}
    \label{fig:ParticleNumberNonConserved}
\end{figure}

In Fig.\,\ref{fig:ParticleNumberNonConserved} (upper frame) the particle number \(N\) is plotted in order to illustrate that the model does not conserve particle number so far. The increasing particle number is fixed by a shrinking chemical potential \(\mu (t)\), lower frame. This also shows why for equal initial and final temperature no increase in condensed particles can be achieved: The particles that were cut off by the truncation must be replaced by the ones at lower energies. Because the equilibrium distribution is equal to the initial distribution without the cutoff, no further particles are available that can be redistributed. 

Nevertheless, the time-dependent chemical potential can be used to impose particle-number conservation: When inserting
\begin{figure}[bt]
    \centering
    \includegraphics[width = 0.50\textwidth]{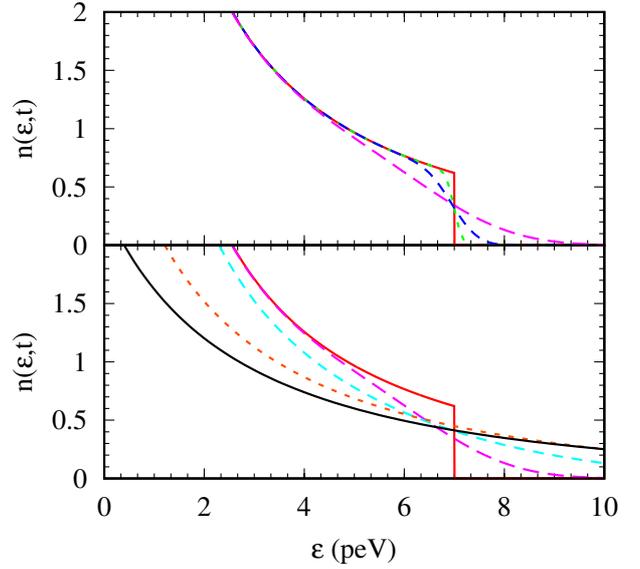}
    \caption{Time-dependent occupation-number distribution with particle-number conserving \(\mu(t)\) describing the equilibration for $T_\text{i} = T_\text{f} = 8$\,peV from an initial distribution at $t=0$ (solid, red) to a Bose-Einstein distribution (solid, black). In the upper part $t=\,$\(\SI{1}{\micro \second}\) (short-dashed, green), \(\SI{10}{\micro \second}\) (medium-dashed, blue) and \(\SI{100}{\micro\second}\) (long-dashed, purple) are displayed;   in the lower part $t=\,$\(\SI{100}{\micro\second}\) (long-dashed, purple), \(\SI{1}{\milli \second}\) (medium-dashed, turquoise) and \(\SI{10}{\milli \second}\) (short-dashed, orange) are shown together with the equilibrium distribution.}
    \label{fig:EquilibrationFixedMu}
\end{figure}
it into the solution Eq.\,\eqref{eq:particledistributionfixedmu}, a particle-number conserving equilibration is obtained, see Fig.\,\ref{fig:EquilibrationFixedMu}. Indeed the number of condensed particles decreases for equal initial and final temperatures, as is obvious from the shift to lower energies of the final Bose-Einstein distribution: This produces a lower intersection of the distribution with the \(y\)-axis that represents the decreasing number of condensed particles.
\subsection{Evaluation for \(T_{\mathrm{i}} \neq T_{\mathrm{f}}\)}
\label{subsec:EvaluationforTineqTf}
\begin{figure}[b!]
    \centering
    \includegraphics[width = 0.50\textwidth]{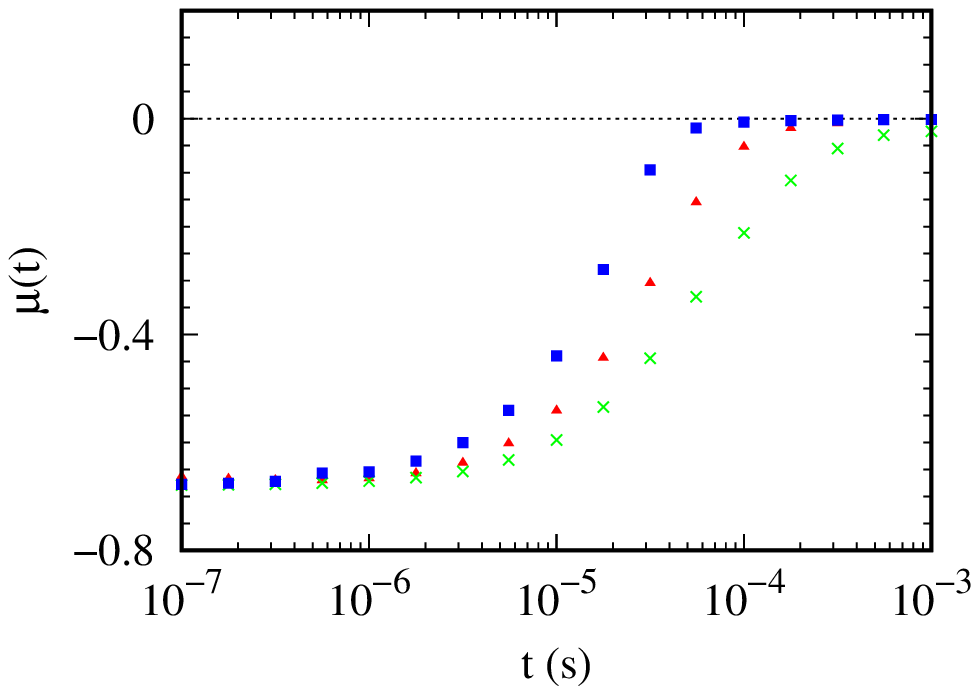}
    \caption{Time-dependent chemical potential for initial temperatures $T_\text{i}$ = \(\SI{30}{\pico \electronvolt}\) (\(\Box\), blue), \(\SI{20}{\pico \electronvolt}\) (\(\bigtriangleup\), red) and \(\SI{15}{\pico \electronvolt}\) (\(\times\), green).}
    \label{fig:MuArbTemp}
\end{figure}

Next we investigate the equilibration with arbitrary initial temperature which differs from the final equilibrium temperature. Thus we can account schematically for evaporative cooling,
and for a redistribution of particles from the thermal cloud into the condensate. Again it can be checked that the solution with fixed chemical potential does not obey particle-number conservation, such that a time-dependent chemical potential becomes necessary. The total particle number at 
$t = 0$ is calculated and set equal to the expression for the particle number at any time. By using numerical methods in Mathematica, we solve this equation and determine the time-dependent chemical potentials. To avoid excessive computing times, the equation is simplified in three ways:
\begin{itemize}
    \item The integration in Eq.\,\eqref{eq:thermalBosons} is not computed up to infinity, but terminated at \(\SI{100}{\pico \electronvolt}\). The neglected part of the integral is vanishingly small because the particle distribution converges to zero for high energies.
    \item The infinite sum in the particle distribution Eq.\,\eqref{eq:Zarbtemp} is confined to the first \(100\) terms. Due to the convergence of the sum, higher order terms are negligibly small and can be omitted.
    \item The chemical potential is calculated pointwise and not continuously.
\end{itemize}
With these simplifications, the chemical potentials for different initial temperatures are plotted in Fig.\,\ref{fig:MuArbTemp}.
For all initial temperatures $T_\text{i} > T_\text{f}$, the chemical potentials are seen to rise with time toward $\mu = 0$ up to the millisecond scale as a consequence of the cooling process, which redistributes atoms from the thermal cloud into the condensate. The larger the difference
between the initial and the final temperature, the faster the rise occurs. It balances the shrinking distribution observed in Fig.\,\ref{fig:Temp20SmallTimes} at $t \gtrsim\,$\(\SI{100}{\micro \second}\). In contrast, at very short times below \(\SI{10}{\micro \second}\) the chemical potential stays nearly constant because of the almost particle-number conserving short-range equilibration around the truncation energy. 
\begin{figure}[t!]
    \centering
    \includegraphics[width = 0.50\textwidth]{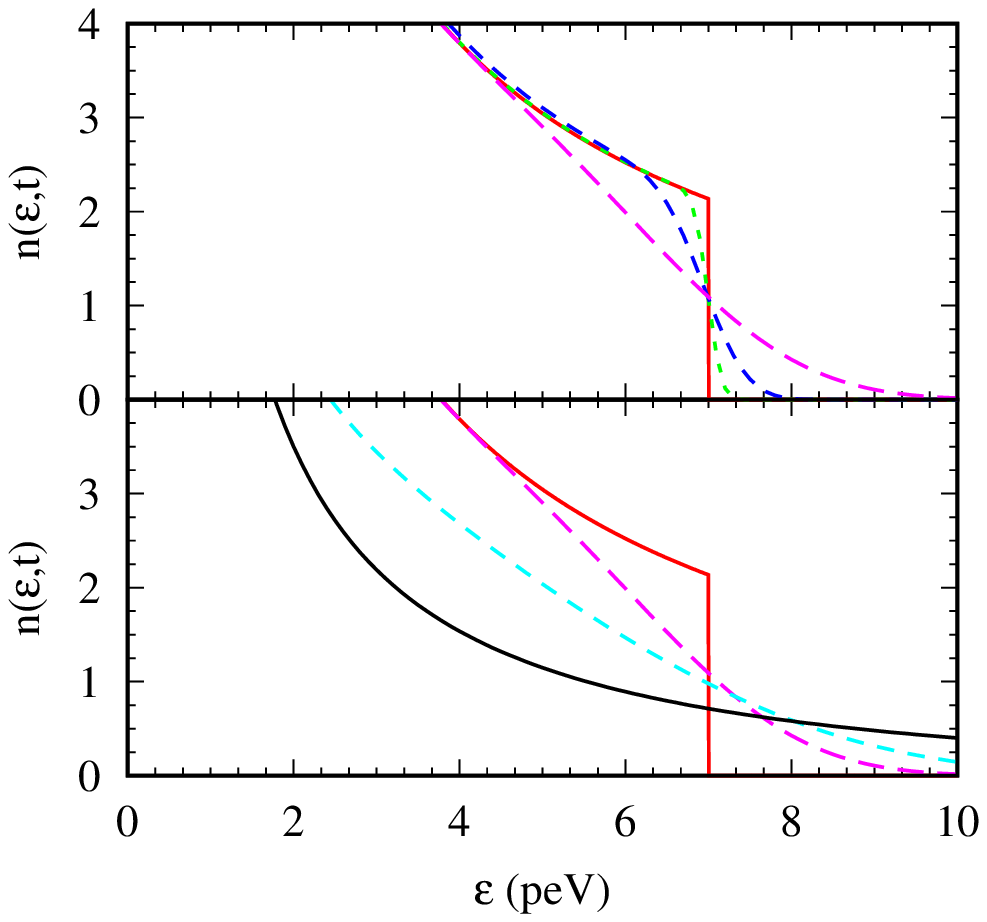}
    \caption{Time-dependent occupation-number distribution for particle-number conserving \(\mu(t)\) describing the equilibration from an initial distribution at $t = 0$ and \(T_\text{i}=\SI{20}{\pico \electronvolt}\) (solid, red) to a Bose-Einstein distribution at \(T_\text{f}=\SI{8}{\pico \electronvolt}\) (solid, black). In the upper part $t=\,$\(\SI{1}{\micro \second}\) (short-dashed, green), \(\SI{10}{\micro \second}\) (medium-dashed, blue) and \(\SI{100}{\micro\second}\) (long-dashed, purple) are displayed;  in the lower part $t=\,$\(\SI{100}{\micro\second}\) (long-dashed, purple) and \(\SI{316}{\micro \second}\) (medium-dashed, turquoise) are shown together with the equilibrium distribution (solid, black).}
    \label{fig:Temp20TimesCorrected}
\end{figure}


Fig.\,\ref{fig:Temp20TimesCorrected} displays the equilibration for \(T_{\mathrm{i}} = \SI{20}{\pico \electronvolt}\) as in 
Fig.\,\ref{fig:Temp20SmallTimes}, but now with time-dependent chemical potentials to conserve the particle number. Hence, the equilibration as modeled through the solution of the NBDE with appropriate boundary conditions at the singularity now represents a cooling process with a gain in condensed particles, as is appropriate for the physical situation.

\section{Conclusion and Outlook}
\label{sec:ConclusionandOutlook}
We have calculated exact analytic solutions of  the nonlinear boson diffusion equation that converge to a Bose-Einstein equilibrium solution. Compared to the solutions obtained in Refs.\,\cite{gw18a} and \cite{gw19} with initial conditions that excluded the singularity and hence, did not include boundary conditions, this is a substantial improvement of the model. Analytical solutions simplify the analysis of the time-dependent equilibration and facilitate further discussions of particle-number conservation.

The bosonic equilibration processes discussed in this work share some common features, like the separation into two time regions. For short times, a local equilibration near the cutoff due to evaporative cooling occurs that is essentially the same for all the different solutions that were investigated. 
Even similarities to the equilibration in fermionic systems were recognized. At large times, however, differences appear depending, in particular, on the treatment of the chemical potential. Nevertheless, a global equilibrium distribution is always approached. In order to determine the boundary between both time regions, we have chosen the time at which the inflection points vanish.

Concerning particle-number conservation, time-dependent chemical potentials have been obtained that imply conserved particle number. As noticed in Sec.\,\ref{subsec:CalcoftheParticleNumber}, the corresponding solutions of the NBDE are then only approximately valid due to the neglect of the time-derivative contribution of the chemical potential. This problem needs further investigations using both numerical and analytical tools: One could try to solve the nonlinear diffusion equation with an ansatz that already contains a time-dependent chemical potential. This leads to a system of coupled equations containing the nonlinear diffusion equation and the total particle number. However, solving a nonlinear partial differential equation and an integral equation in a coupled system will require much more effort, and is unlikely to allow for exact analytical solutions.

Various other considerations are conceivable. For instance, the critical temperature could be determined, at which the cooling starts to cause the rise of the condensed particles.
The resulting value can be compared to data, and to the equilibrium-statistical prediction. 

A related model has been developed for fermionic systems \cite{gw82,gw18}. The nonlinear fermion  diffusion equation can be derived in a similar way as the one for bosons.
The static solution of this equation is the Fermi-Dirac distribution, it represents the fermionic equilibrium occupation-number distribution. For this equation an analytic solution can be found as well, and the time-dependent equilibration has been investigated for several initial distributions in 
Ref.\,\cite{bgw19}. 

Due to the lack of a singularity in the fermionic case, no boundary condition is required. Interestingly, for a constant density of states the solutions obey particle-number conservation even at high relativistic energies, provided the creation of anti-particles is also taken into account. 
Through the similarity of both models, comparisons \cite{gw19} to the results for fermions have been instrumental in our improvement of the bosonic model. In particular, having found converging solutions for the nonlinear fermion diffusion equation was a key for the assertion that the bosonic equation should have equilibrating solutions as well  -- which indeed has turned out to be the case.

Another promising field of research would be the microscopic analysis of the transport coefficients \(D\) and \(\varv\) that were assumed to be constant in this work. This conjecture entails not only the correct equilibrium solution, but also exact time-dependent solutions. A microscopic derivation of the transport coefficients in a relativistic many-body theory will certainly be helpful to motivate this step,
provided it results in sufficiently weak dependencies on energy and time.

A model that aims to fully account for the redistribution of particles into the condensate would, moreover, have to include a quantum-mechanical description \cite{gro61,pit61} with the trapping potential, since the analytic results in this work were essentially
derived within a semiclassical limit, and without a confining potential. 

Despite such issues, we have solved analytically a kinetic model that achieves three basic aims: The solutions converge to equilibrium solutions, which have been generalized for arbitrary initial temperatures that differ from the equilibrium temperature. Therefore, cooling is accounted for as well, and our approach can be extended to sequential cooling steps as in the experiments. Finally, particle-number conservation has also been introduced into the model. Hence, our approach is a good tool for further investigations of equilibration processes in cold bosonic gases, eventually providing comparisons to data for these atomic systems at low temperatures.\\

\bf{Acknowledgements}\rm\\

One of us thanks J. H\"olck (Heidelberg University) for discussions about solution methods for the nonlinear boson diffusion equation with boundary conditions, and A. Simon for a detailed comparison of our analytic NBDE solutions with Matlab-results. NR acknowledges support of the Studienstiftung, Germany.

\bibliographystyle{elsarticle-num}
\bibliography{gw_19.bib}
\end{document}